\documentclass[prd,twocolumn,twoside,preprintnumbers,superscriptaddress,nofootinbib]{revtex4}

\usepackage{amsmath,slashed}
\usepackage{graphicx,graphics,color}
\usepackage{xcolor}
\usepackage{dcolumn}
\usepackage[hyperfootnotes=false]{hyperref}
\usepackage{xspace}
\usepackage{enumerate}
\usepackage{varwidth}
\usepackage{caption}
\usepackage{subcaption}
\usepackage{comment}

\usepackage{amssymb}
\usepackage{mathtools}
\usepackage{dsfont}
\usepackage{multirow}

\newcommand{\beq}{\begin{equation}}
\newcommand{\eeq}{\end{equation}}

\newcommand{\bea}{\begin{eqnarray}}
\newcommand{\eea}{\end{eqnarray}}

\allowdisplaybreaks[1]

\begin{document}

\preprint{LA-UR-26-27228}

\title{Radiative corrections to neutron $\beta$ decay with explicit $\Delta (1232)$ degrees of freedom}

\author{Lemonia Gialidi}
\affiliation{Institute for Theoretical Physics Amsterdam and Delta Institute for Theoretical
Physics, University of Amsterdam, Science Park 904, 1098 XH Amsterdam, The Netherlands}
\affiliation{Nikhef, Theory Group, Science Park 105, 1098 XG, Amsterdam, The Netherlands}
\author{Emanuele Mereghetti }
\affiliation{Los Alamos National Laboratory, Theoretical Division T-2, Los Alamos, NM 87545, USA}
\author{Jordy de Vries}
\affiliation{Institute for Theoretical Physics Amsterdam and Delta Institute for Theoretical
Physics, University of Amsterdam, Science Park 904, 1098 XH Amsterdam, The Netherlands}
\affiliation{Nikhef, Theory Group, Science Park 105, 1098 XG, Amsterdam, The Netherlands}

\begin{abstract}

We study electromagnetic radiative corrections to neutron $\beta$ decay within chiral perturbation theory extended to include the $\Delta(1232)$ resonance as an explicit degree of freedom. Building on a previous analysis performed in the $\Delta$-less theory, we identify and compute additional loop contributions with
intermediate $\Delta$ states at leading and next-to-leading order in the small-scale expansion, and match the results to the pionless effective field theory. We find that the convergence of the chiral expansion significantly improved in the $\Delta$-full calculation. The electromagnetic shift in the axial coupling is reduced by $25\%$ compared to the $\Delta$-less result. The correction remains at the percent level and is phenomenologically significant for the comparison between experimental extractions and lattice-QCD determinations of the nucleon axial charge. We also obtain a
new $\Delta$-induced contribution to the weak magnetism term, which is an order of magnitude smaller than the pion-loop result and negligible for upcoming experiments.

\end{abstract}

\maketitle

\section{Introduction}
Weak decays of strongly-interacting systems, like mesons, neutrons, and atomic nuclei, provide high-precision tests of the Standard Model (SM) and the ideal setting to perform sensitive searches for physics beyond-the-SM (BSM). An important example is the role of $\beta$ decays in the extraction of the Cabibbo-Kobayashi-Maskawa (CKM) matrix element $V_{ud}$. Following recent studies of $\beta$ decay rates at the sub-percent level \cite{Seng:2018yzq,Seng:2018qru,Czarnecki:2019mwq,Shiells:2020fqp,Hardy:2020qwl}, there is a $\sim 3\sigma$ tension with the SM prediction based on the unitarity of the CKM matrix \cite{Hardy:2020qwl,ParticleDataGroup:2020ssz}. In addition, decays of particles with spin, such as the neutron, can give access to a richer set of observables, such as the ratio of axial to vector couplings, $g_A/g_V$, which can be a sensitive probe of BSM physics at the multi-TeV scale. A particularly sensitive test is provided by the comparison between the experimentally extracted axial coupling $g_A$ and its determination from lattice QCD \cite{Bhattacharya:2011qm,Alioli:2017ces,Chang:2018uxx,Jang:2023zts,Alexandrou:2023qbg,Hall:2025ytt}, which probes possible right-handed charged currents. Compared to nuclei, the neutron can also be more advantageous, as it does not depend on nuclear structure uncertainties, and lattice calculations are more precise.

Experimental high-precision measurements of the neutron decay lifetime $\tau_n$ and the axial-to-vector coupling ratio are reaching $10^{-4}$ precision \cite{Musedinovic:2024gms,UCNt:2021pcg,Markisch:2018ndu,ParticleDataGroup:2022pth}. To identify BSM effects, we need to reach the same precision in the SM predictions. In particular, this requires control over recoil and radiative corrections, spurring a lot of recent activity to improve the theoretical description of both Fermi (vector) 
\cite{Seng:2018yzq, Seng:2018qru,Czarnecki:2019mwq,Shiells:2020fqp,Cirigliano:2023fnz}
, and Gamow-Teller (axial)  \cite{Gorchtein:2021fce,Hayen:2019nic,Cirigliano:2022hob,Cirigliano:2024nfi,Tomalak:2026wks} transitions. 

From an effective field theory (EFT) point of view, neutron $\beta$ decay involves a wide range of scales from possible multi-TeV BSM scales to the proton-neutron mass splitting, a few MeV.  This separation of scales ensures that neutron $\beta$ decay is an ideal candidate for an EFT description. In the pionless EFT of Ref.~\cite{Ando:2004rk}, where the relevant degrees of freedom are nucleons, leptons, and photons, we can treat recoil effects and long-distance QED corrections in a systematic way. However, hadronic structure effects associated with the pion scale are encoded in low-energy constants and cannot be computed explicitly within this framework. To access these contributions, we must instead work within chiral perturbation theory \cite{Bernard:1995dp}, where pions appear as dynamical degrees of freedom.

Radiative corrections to neutron $\beta$ decay induced by pion loops were studied in Ref. \cite{Cirigliano:2022hob}, where previously overlooked  electromagnetic corrections to the axial coupling $g_A$ were computed. These corrections were found to induce a percent-level shift in $g_A$, significantly larger than earlier estimates, highlighting the importance of long-distance hadronic structure effects. 

The analysis of Ref.~\cite{Cirigliano:2022hob} was performed in a theory without explicit $\Delta (1232)$ degrees of freedom and the dominant radiative contribution arises from diagrams involving the low-energy coefficients (LECs) $c_3$ and $c_4$. In a $\Delta$-less theory, these LECs are rather large and, for example, lead to a poorly convergent chiral expansion for the axial coupling $g_A$ \cite{Chang:2018uxx,Hall:2025ytt,Ekstrom:2025ost}.
Similarly, formally next-to-leading-order (NLO) radiative corrections to neutron $\beta$ decay were found to be numerically larger than lower order terms \cite{Cirigliano:2022hob}. This suggests that the convergence of the $\Delta$-less chiral expansion could potentially be improved by treating the nearby $\Delta(1232)$ resonance, with a mass only $300$ MeV larger than the nucleon,  as an explicit degree of freedom, rather than integrating it out into the low-energy constants (LECs)~\cite{Pascalutsa:2002pi,Hemmert:1997ye}.
The main goal of this work is to investigate the role of explicit $\Delta$ dynamics on the pionic radiative corrections of Ref.~\cite{Cirigliano:2022hob} and assess if they improve the chiral convergence and affect the quantitative estimate of the radiative corrections to $g_A$ and neutron $\beta$ decay in general.  

The paper is organized as follows. 
In Sec.~\ref{sec:framework} we review the EFT framework for neutron $\beta$ decay at $\mathcal O(\alpha)$. In Sec.~\ref{sec:deltaloops} we calculate loop corrections with intermediate $\Delta$ baryons. In Sec.~\ref{sec:Numerics} we discuss the numerical impact of including the $\Delta$ as an explicit degree of freedom, and we conclude in Sec.~\ref{sec:discussion}.

\section{Effective field theory framework}
\label{sec:framework}
\subsection{Chiral Lagrangian with  $\Delta$ degrees of freedom}
We work within chiral perturbation theory ($\chi$PT), which provides a systematic description of low-energy QCD in terms of pions and baryons \cite{Gasser:1983yg,
Bernard:1995dp}. In the baryon sector, the heavy-baryon formulation allows for a consistent power counting and has been widely used in the study of low-energy processes \cite{Jenkins:1990jv,Bernard:1995dp,Meissner:1997ii,Muller:1999ww,Gasser:2002am}. 

As discussed in the introduction, we extend the chiral approach by including the lightest baryonic resonance, namely the $\Delta(1232)$, as a dynamical degree of freedom~\cite{Jenkins:1991es,Hemmert:1997ye}. This is particularly relevant given the relatively small mass splitting between the nucleon and the $\Delta$, which can influence observables even at low energies, thereby improving the description of nucleon structure observables and reorganizing the chiral expansion. The four physical states $\Delta^{++}$, $\Delta^+$, $\Delta^0$, and $\Delta^-$ are collected into the spin-$3/2$ isospin multiplet $T^\mu_i$, following the formulation of Ref.~\cite{Hemmert:1997ye}.
The heavy-baryon propagator of the $\Delta$ field is given by~\cite{Jenkins:1991es}
\begin{equation}
D_{\mu\nu}^{ij}(p)
= \frac{i\, P^{3/2}_{\mu\nu}\, \xi_{3/2}^{ij}}{v\!\cdot\!p - \Delta+i0},
\label{eq:DeltaProp}
\end{equation}
where the spin-$3/2$ projector is given by
$P^{3/2}_{\mu\nu} =  v_\mu v_\nu -g_{\mu\nu} -\frac{4}{d-1} S_\mu S_\nu$.

At leading order (LO), the  Lagrangian is given by \cite{Siemens:2020vop}
\begin{align}
\mathcal{L}^{(1)}_{\pi N (\Delta)} &= 
\bar{N} \, i v \cdot D \, N 
+ g_A \, \bar{N} S \cdot u \, N \nonumber \\
&\quad - \bar{T}^\mu_i \left( i v \cdot D^{ij} - \Delta \, \delta^{ij} + g_1 \, S \cdot u^{ij} \right) T^j_\mu \nonumber \\
&\quad + h_A \left( \bar{T}^\mu_i w^i_\mu N + \bar{N} w^{i\dagger}_\mu T^\mu_i \right),
\end{align}
where $N$ denotes the nucleon doublet and $T^\mu_i$ the heavy-baryon field describing the $\Delta$ resonance, with isospin index $i$. 
The superscript $(1)$ denotes the chiral index of the Lagrangian, which counts the powers of momentum or pion mass in the operators.  
We work in the nucleon rest frame with velocity $v^\mu = (1,\mathbf{0})$ and spin operator $S^\mu = (0,\boldsymbol{\sigma}/2)$. The quantity $\Delta = m_\Delta - m_N$ denotes the $\Delta$-nucleon mass splitting. The couplings $g_A$, $g_1$, and $h_A$ correspond to the nucleon, $\Delta$, and the nucleon-$\Delta$ axial couplings, respectively.

The nucleon and $\Delta$ covariant derivatives are defined as
\begin{align}
D_\mu N &= (\partial_\mu + \Gamma_\mu) N, \nonumber\\[4pt]
D_\mu^{ij} T_{\nu j} &= 
\Big( \partial_\mu \delta^{ij} + \Gamma_\mu \delta^{ij} 
      - i\, \epsilon^{ijk} \text{Tr}(\tau^k \Gamma_\mu) \Big) T_{\nu j},
\label{eq:covariantD}
\end{align}
while $\Gamma_\mu$ is given by
\begin{equation}
\Gamma_\mu = 
\frac{1}{2}\left[
u^\dagger \big( \partial_\mu - i(l_\mu + {3}\hat{l}_\mu) \big) u 
+ u \big( \partial_\mu - i(r_\mu + { 3} \hat{r}_\mu) \big) u^\dagger
\right],
\label{eq:Gamma}
\end{equation}
and $u_\mu$ is defined by
\begin{equation}
u_\mu = -i\left[
u^\dagger \big( \partial_\mu - i(l_\mu + \hat{l}_\mu) \big) u
- u \big( \partial_\mu - i(r_\mu + \hat{r}_\mu) \big) u^\dagger
\right]
\label{eq:uMu}
\end{equation}
where the matrix $u$ encodes the pion fields
\begin{equation}
    U = u^2 = \exp\left( \frac{i \pi^a \tau^a}{F_\pi} \right)\,,
    \label{eq:U}
\end{equation}
where $F_\pi\simeq 92.2$ MeV is the pion decay constant. 
For the $\Delta$ interactions, we introduce
\begin{equation}
w^i_\mu = \frac{1}{2} \mathrm{Tr}(\tau^i u_\mu), 
\qquad
u^{ij}_\mu = \xi^{3/2}_{il} \, u_\mu \, \xi^{3/2}_{lj},
\end{equation}
where $\xi^{3/2}_{ij}= \frac{2}{3} \delta_{i j} - \frac{i}{3} \varepsilon_{i j k} \tau^k$ projects onto the isospin-$3/2$ sector.

The coupling to external electromagnetic and weak fields is implemented through left- and right-handed sources $l_\mu$ and $r_\mu$,
\begin{eqnarray}
    l_\mu &=& \frac{e}{2} A_\mu \tau_3 - 2 \sqrt{2} G_F V_{ud} \, \bar{e}_L \gamma_\mu \nu_L \, \tau_+ + \mathrm{h.c.}\,,\nonumber\\
r_\mu &=& \frac{e}{2} A_\mu \tau_3,\qquad \hat l_\mu =\hat r_\mu = \frac{e}{6} A_\mu.
\end{eqnarray}
We define the leptonic amplitude
\begin{equation}
\bar{l}_\mu = -2 \sqrt{2} G_F V_{ud} \, \bar{u}(p_e)\gamma_\mu P_L\;v_{\nu}(p_\nu),
\end{equation}
such that the tree-level neutron decay amplitude can be written as
\begin{equation}
\mathcal{A}_{\text{tree}} = \frac{\bar{l}_\mu}{2} \left( v^\mu - 2 g_A S^\mu \right).
\end{equation}

This structure makes it explicit that corrections to the vector and axial currents can be identified separately. 
In what follows, we will compute loop corrections involving pion and $\Delta$ intermediate states and extract their contributions to the vector and axial structures of the amplitude. We are particularly interested in capturing the electromagnetic corrections arising from the pion mass splitting as these dominated the corrections identified in Ref.~\cite{Cirigliano:2022hob}. At leading order in the electromagnetic interaction, the pion mass splitting is described by \footnote{We note that the overall sign of this term was given incorrectly in~\cite{Cirigliano:2022hob}, here we give the corrected sign.}
\begin{equation}
    \mathcal{L}_\pi^{(e^2 p^0)}=-2 e^2 F_\pi^2 Z_\pi \pi^+ \pi^-+ \mathcal{O}(\pi^4) ,
\end{equation}
where $Z_\pi$ is connected to the pion mass splitting $\delta m_\pi^2=m^2_{\pi^\pm}-m^2_{\pi^0}=2 e^2 F_\pi^2 Z_\pi$.

At next-to-leading order (NLO),  the $\pi N$ Lagrangian contains the usual subleading interactions proportional to the LECs $c_3$ and $c_4$, whose explicit expression can be found in Ref.~\cite{Bernard:1995dp}. At the same order, the relevant part of the effective $N \Delta$ Lagrangian is \cite{Hemmert:1997ye}
\begin{equation}
\begin{split}
    \mathcal{L}^{(2)}_{ N \Delta } &=\, \bar T_i^\mu\,  \bigg[\frac{i\,b_1}{2m_N} f_{+\mu\nu}^i\, S^\nu +(b_3+b_6)\, i w_{\mu\nu}^i v^\nu \\ &+ b_4\, w_\mu^i S\cdot u  + b_5\, u_\mu S\cdot w^i \bigg]N + \text{h.c.}\; ,
\end{split}
\end{equation}
where $b_i$ are the NLO LECs, $w_{\mu\nu}^i=\frac{1}{2}\text{Tr} \left[\tau^i [D_\mu,u_\nu]\right]$ and $f^+_{\mu\nu} = u F^L_{\mu\nu} u^\dagger + u^\dagger F^R_{\mu\nu} u\equiv\tau^if^i_{+\mu\nu} $.
For pure electromagnetic interactions and no weak axial fields, we define
\begin{equation}
    F^L_{\mu\nu} = F^R_{\mu\nu} = -e\,Q\,F_{\mu\nu},\qquad
F_{\mu\nu} = \partial_\mu A_\nu-\partial_\nu A_\mu.
\end{equation}
The operator proportional to the dimensionless LEC $b_1$ induces the magnetic dipole (M1) transition between the nucleons and the $\Delta$. The operators proportional to $b_3,b_6$ here can be removed by using field redefinitions, and therefore only the dimensionful $b_4,b_5$ generate independent NLO contributions that are relevant for the present work. By dimensional analysis, one expects $b_{4,5} =\mathcal O(\Lambda_\chi^{-1})$, where $\Lambda_\chi$ is the $\chi$PT breakdown scale. Additional recoil interactions appear at this order with LECs proportional to $1/m_N$. We neglect these terms following the power counting assignments of Ref.~\cite{Epelbaum:2008ga}.

At $\mathcal O(e^2G_Fp^0)$, the loop contributions are accompanied by local counterterms. The local operators that contribute to the axial part of the neutron decay amplitude have the same structure as the leading axial current. We collect the relevant combination of the renormalized LECs in the effective coefficient $\hat{C}_A(\mu)$ and define its contribution to the amplitude as
\begin{equation}
    \mathcal A_{\rm ct}=\frac{\bar l_\mu}{2}\left(-2g_A S^\mu\right)\frac{\alpha}{2\pi}\hat C_A(\mu)
\end{equation}
The scale dependence of $\hat C_A(\mu)$ cancels that of the loop contributions, so the complete radiative correction of the axial current is renormalization-scale independent. The full chiral-invariant operator construction and the expression of $\hat C_A$ in terms of the underlying LECs can be found in Refs.~\cite{Cirigliano:2022hob,Cirigliano:2024nfi}. 
As the nucleon-$\Delta$ mass splitting is a chiral singlet,
including explicit $\Delta$ degrees of freedom does not introduce a new pionless axial structure or new counterterms, but instead modifies the matching and renormalization of $\hat C_A(\mu)$.
In the $\Delta$-full theory, we can therefore think of the counterterms as functions of $\mu/\Delta$ and of having an expansion in powers of $\Delta/\Lambda_\chi$ \cite{Tiburzi:2005na}
\begin{equation}\label{eq:ctexp}
 \hat{C}_A = \hat C_A^{(0)} + \frac{\Delta}{\Lambda_\chi} \hat C_A^{(1)} + \ldots. 
\end{equation}
As we discuss in what follows, we can leverage this structure to renormalize LO and NLO  diagrams with intermediate $\Delta$, and to absorb in $\hat{C}_A$ terms that would naively break decoupling.

\subsection{Power counting and decoupling}

In the presence of explicit $\Delta$ degrees of freedom, we adopt the small-scale expansion (SSE)~\cite{Hemmert:1997ye}, 
\begin{equation}
\epsilon_\chi \sim \left\{ \frac{Q}{\Lambda_\chi}, \frac{m_\pi}{\Lambda_\chi}, \frac{\Delta}{\Lambda_\chi} \right\},
\end{equation}
with $\Lambda_\chi \sim  4\pi F_\pi \sim m_N$ the chiral-symmetry-breaking scale that numerically is close to the nucleon mass. In the SSE, the $\Delta$-nucleon mass splitting is considered a light scale, $\Delta \sim m_\pi \ll \Lambda_\chi$, similar to the pion mass.

Within this framework, we can systematically organize loop contributions involving intermediate $\Delta$ states \cite{Hemmert:1997ye, Siemens:2020vop}. We will see that loop diagrams with pion and $\Delta$ intermediate states contribute already at leading order. Before calculating these loops, there is a technical complication that should be addressed. Loop integrals with explicit $\Delta$ give contributions that scale logarithmically or even with positive powers of the mass splitting $\Delta$, which do not vanish in the limit $\Delta \to \infty$. They therefore violate the expected decoupling of the $\Delta$ as a heavy degree of freedom. Following Ref.~\cite{Siemens:2020vop}, we deal with this by separating the power counting of diagrams from the renormalization procedure. While we use the SSE for power counting to identify the relevant contributions at a given order, in the renormalization procedure, we effectively treat $\Delta$ as a heavy scale, making sure that terms analytic in the low-energy scales, including those proportional to positive powers of $\Delta$, are absorbed into LECs.
After this procedure, the remaining contributions are suppressed by inverse powers of $\Delta$ and vanish in the decoupling limit. This separation will be implemented explicitly in Sec.~\ref{sec:deltaloops}, where we isolate the contributions relevant for the radiative corrections. We give an explicit example of this procedure in the appendix. 

This procedure is also employed in the extraction of the $\pi N$ LECs from $\pi N$ scattering in the $\Delta$-full theory ~\cite{Siemens:2020vop}. Using the same renormalization scheme here ensures that the values of $c_3$ and $c_4$ used in our numerical analysis are determined consistently with the present calculation.

\subsection{Matching to the pionless theory}
In neutron $\beta$ decay, the typical external momenta and the charged lepton mass are much smaller than the pion mass
\begin{equation}
q_{\mathrm{ext}},\, m_e \ll m_\pi , \Delta\ll \Lambda_\chi .
\end{equation}
This hierarchy makes it natural to describe the process in terms of a pionless effective field theory, in which pions and $\Delta$ baryons are integrated out. 
At low energies, the non-relativistic effective pionless Lagrangian is given by
\begin{eqnarray}\label{eq:Lpionless}
 \mathcal L_{\slashed \pi} &=& 
 - \sqrt{2}  G_F V_{ud}  \, \bigg[
  \bar e \gamma_\mu P_L \nu_e 
   \bigg(
  \bar N  \left(g_V  v_\mu - 2 g_A  S_\mu \right)  \tau^+ N  
 \nonumber \\ 
 &+ & \frac{i}{2 m_N}  \bar N 
  (v^\mu v^\nu - g^{\mu \nu}  - 2 g_A v^\mu S^\nu) (  \overleftarrow   \partial - \overrightarrow \partial)_\nu  \tau^+ N  
  \bigg)
  \nonumber \\ 
 & +& 
 \frac{i \mu_{\rm weak}}{m_N}    \bar N [S^\mu, S^\nu]  \tau^+ N\,   \partial_\nu \left(  \bar e \gamma_\mu P_L \nu \right)
  \nonumber \\ 
&+ &  \frac{i c_T m_e}{m_N} \bar N \left( S^\mu v^\nu - S^\nu v^\mu\right)  \tau^+ N\,   \left(  \bar e \sigma_{\mu \nu} P_L \nu \right)  + \ldots \bigg]\,. \nonumber \\
\end{eqnarray}
The operators in the pionless Lagrangian are organized in powers of two small parameters 
\begin{enumerate}
    \item $\epsilon_{\text{recoil}}=q_{\mathrm{ext}}/ m_N$ related to small kinetic corrections,
    \item $\epsilon_{\slashed{\pi}}=\{ q_{\mathrm{ext}}/ m_\pi$, $q_{\mathrm{ext}}/\Delta$\}, nucleon structure corrections associated with radiative pion contributions.
\end{enumerate}
We truncated the expression in Eq.~\eqref{eq:Lpionless} to terms of $\mathcal O(\epsilon_\text{recoil})$
and $\mathcal O(\epsilon_\slashed{\pi})$, while $\ldots$ denotes terms suppressed by more power of external momenta.
The LECs in the pionless EFT are obtained by matching to $\chi$PT, and they have an expansion in $\epsilon_\chi$ and in $\alpha$. 
In the next subsection we will review the determination of $g_A$, $c_T$ and $\mu_{\text{weak}}$ in $\chi$PT without $\Delta$ degrees of freedom.

\subsection{Pion-induced radiative corrections}
We start by considering terms in $\mathcal L_\slashed{\pi}$ not suppressed by external momenta, which reduce to the vector and axial couplings $g_V$ and $g_A$.
We can express these couplings in a double expansion in $\epsilon_\chi$ and $\alpha$ as
\begin{equation}\label{eq:expansion}
 g_{V/A}  = g_{V/A}^{(0)} \left[1 + \sum_{n=2}^{\infty} \Delta_{V/A,\chi}^{(n)} +  \frac{\alpha}{2\pi} \sum_{n=0}^{\infty} \Delta^{(n)}_{V/A, \mathrm{em}}+\dots \right].
\end{equation}
Compared to Ref.~\cite{Cirigliano:2022hob}, 
we are neglecting terms proportional to the quark mass splitting, which are small~\cite{Behrends:1960nf,Ademollo:1964sr,Seng:2023jby}.
$g_{V/A}^{(0)}$ denote the vector and axial couplings in the chiral limit. The second term in the square bracket captures corrections that arise in pure QCD in the isospin limit, with 
$\Delta^{(n)}_{V/A, \chi} \sim \mathcal O(\epsilon_\chi^n)$.
In the case of the vector coupling, vector current conservation implies $g_V^{(0)}=1$ and $\Delta^{(n)}_{V, \chi} = 0$. 
For the axial coupling, chiral corrections start at $n=2$. The $n=2,3$ corrections have been computed in both $\Delta$-less and $\Delta$-full $\chi$PT
\cite{Bernard:1992qa,Bernard:1995dp,Kambor:1998pi,Beane:2004rf,Hemmert:2003cb}. In $\Delta$-less $\chi$PT, two-loop corrections at $n=4$ have also been computed \cite{Bernard:2006te, Bernard:2025gto}. 
For our purposes, we can simply identify
\begin{equation}
    g_A^{\text{QCD}} = g_A^{(0)}  \left[1 + \sum_{n=2}^{\infty} \Delta_{A,\chi}^{(n)}  \right], 
\end{equation}
with $g_A^{\text{QCD}}$ the nucleon axial coupling computed in Lattice QCD.
The last term in Eq.~\eqref{eq:expansion} denotes $\mathcal O(\alpha)$ radiative corrections, which can also receive contributions at different chiral orders,
$\Delta^{(n)}_{{\rm em}} \sim \mathcal O(\epsilon_\chi^n)$. The $\ldots$ denote contributions of $\mathcal O(\alpha^2)$ or higher, which will similarly have their own chiral expansion.

To $\mathcal O(\alpha \epsilon_\chi^0)$, the shift to the axial coupling is 
\begin{equation}
 \Delta^{(0)}_{A, \rm em}
 = Z_\pi \left[\frac{1 + 3 g^{ 2}_A}{2}
 \left(  \log \frac{\mu^2}{m_\pi^2} -1 \right) - g_A^{2} \right]+ \hat{C}_{A}(\mu)~.
 \label{eq:dAem0}
\end{equation}
One order higher, at $\mathcal O(\alpha \epsilon_\chi)$, the correction to $g_A$ is
\begin{eqnarray}\label{c4c3}
 \Delta^{(1)}_{A, \rm  em}
  = Z_\pi \,  4\pi m_\pi  \left[c_4 - c_3  + \frac{3}{8 m_N} + \frac{9}{16 m_N} g_A^{2}\right]\,.
\end{eqnarray}
This correction is dominated by the LECs  $c_{3,4}$ and numerically larger than the formally lower order $\Delta^{(0)}_{A, \rm em}$ correction. We get back to this in Sec.~\ref{sec:Numerics}. 

In addition to corrections to $g_A$, pion loops induce radiative corrections to weak magnetism 
\begin{equation}\label{eq:WMPRL}
\mu_{\mathrm{weak}
} -(\mu_p - \mu_n) = -\frac{\alpha Z_\pi}{2\pi}\frac{g_A^2 m_N \pi}{m_\pi}\,,
\end{equation}
where $\mu_{n,p}$ are the nucleon magnetic moments. This shift leads to $\mathcal O(10^{-4})$ corrections to the $\beta$-$\nu$ angular correlation (the $a$ coefficient) and the $\beta$-asymmetry (the  $A$ coefficient), which are comparable to the anticipated precision of future experiments \cite{Cirigliano:2019wao}. The same loops also induce a nonzero tensor coupling
\begin{equation}\label{eq:cTPRL}
c_T = \frac{\alpha}{2\pi}\frac{g_A m_N \pi}{3 m_\pi}\,.
\end{equation}
$c_T$ contributes to the Fierz interference term $b$,
but the correction is too small to be observable with the current and expected experimental sensitivity. 

Below we compute how loops with explicit $\Delta$ baryons affect the correction to $g_A$ and $\mu_{\mathrm{weak}}$ ($c_T$ corrections are too small).  Note that in the $\Delta$-full theory, the diagrams discussed in Ref.~\cite{Cirigliano:2022hob} remain present, and therefore the corrections discussed above are included in our calculation. However, the dominant radiative correction is the term proportional to $c_4-c_3$ in Eq.~\eqref{c4c3} which takes on a different (smaller) numerical value in the $\Delta$-full theory ~\cite{Siemens:2016jwj}.

\section{Loop contributions with intermediate $\Delta$} \label{sec:deltaloops}
\begin{figure*}[t]
    \centering
    \begin{subfigure}[b]{0.2\textwidth}
         \centering
         \includegraphics[width=\textwidth]{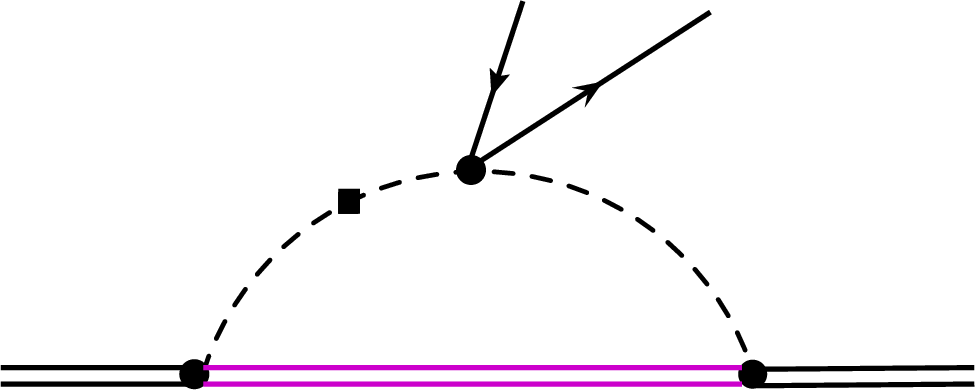}
         \caption{}
         \label{fig:pion1}
    \end{subfigure}
    \hspace{0.02\textwidth}
    \begin{subfigure}[b]{0.2\textwidth}
         \centering
         \includegraphics[width=\textwidth]{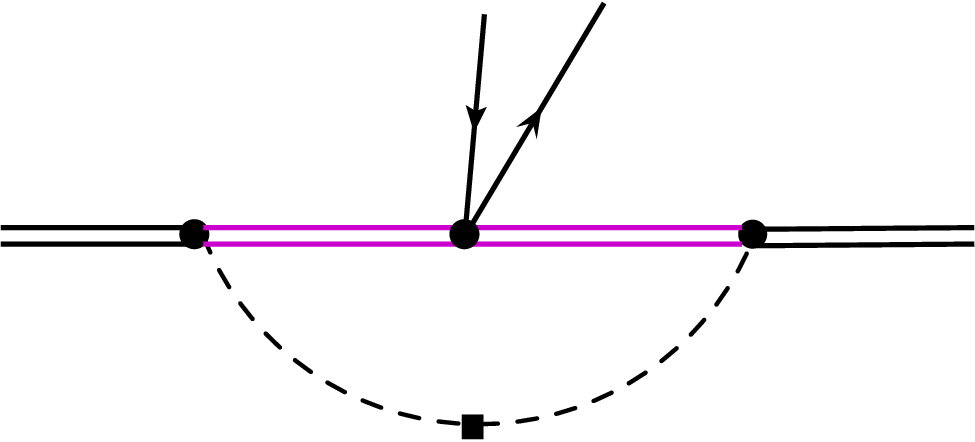}
         \caption{}
         \label{fig:pion2}
    \end{subfigure}
    \hspace{0.02\textwidth}
    \begin{subfigure}[b]{0.2\textwidth}
         \centering
         \includegraphics[width=\textwidth]{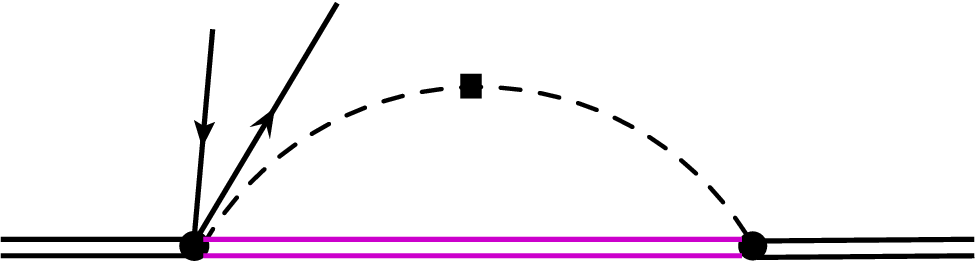}
         \caption{}
         \label{fig:pion3} 
    \end{subfigure}
    \hspace{0.02\textwidth}
    \begin{subfigure}[b]{0.2\textwidth}
         \centering
         \includegraphics[width=\textwidth]{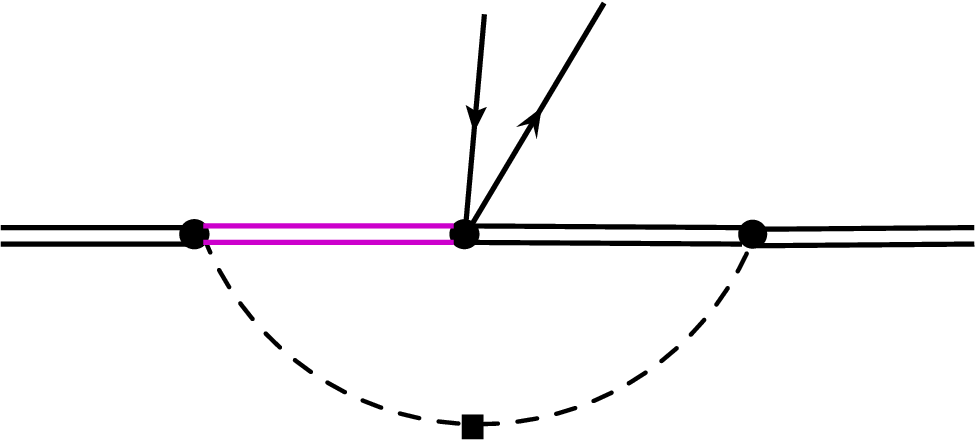}
         \caption{}
         \label{fig:pion4} 
    \end{subfigure}
    \hspace{0.02\textwidth}
    \begin{subfigure}[b]{0.2\textwidth}
         \centering
         \includegraphics[width=\textwidth]{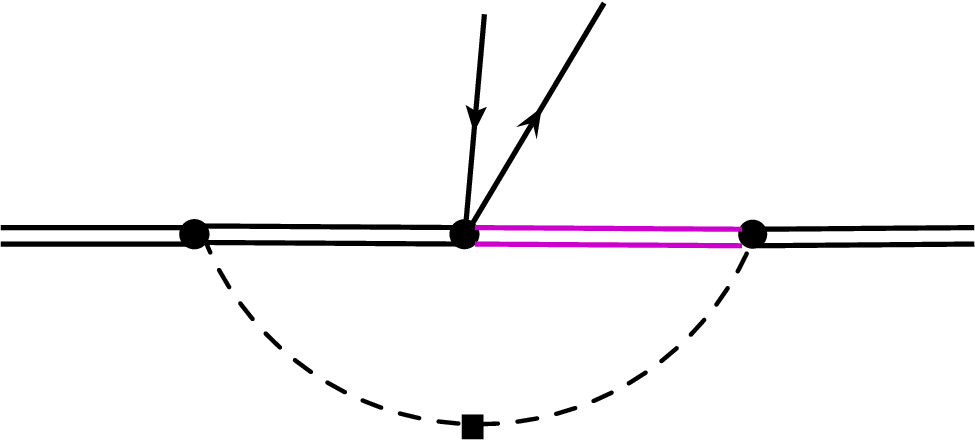}
         \caption{}
         \label{fig:pion5} 
    \end{subfigure}
    \hspace{0.02\textwidth}
    \begin{subfigure}[b]{0.2\textwidth}
         \centering
         \includegraphics[width=\textwidth]{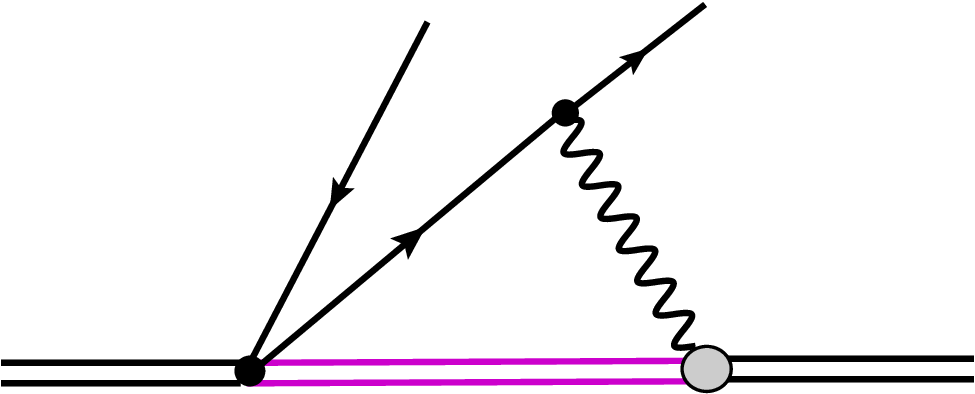}
         \caption{}
         \label{fig:vertex1}
    \end{subfigure}
    \hspace{0.02\textwidth}
    \begin{subfigure}[b]{0.22\textwidth}
         \centering
         \includegraphics[width=\textwidth]{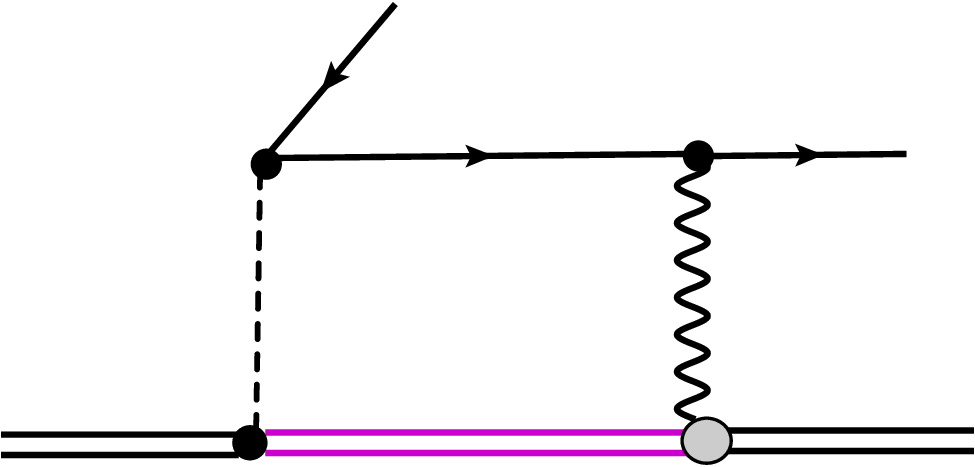}
         \caption{}
         \label{fig:box}
    \end{subfigure}
    \hspace{0.02\textwidth}
    \begin{subfigure}[b]{0.22\textwidth}
         \centering
         \includegraphics[width=\textwidth]{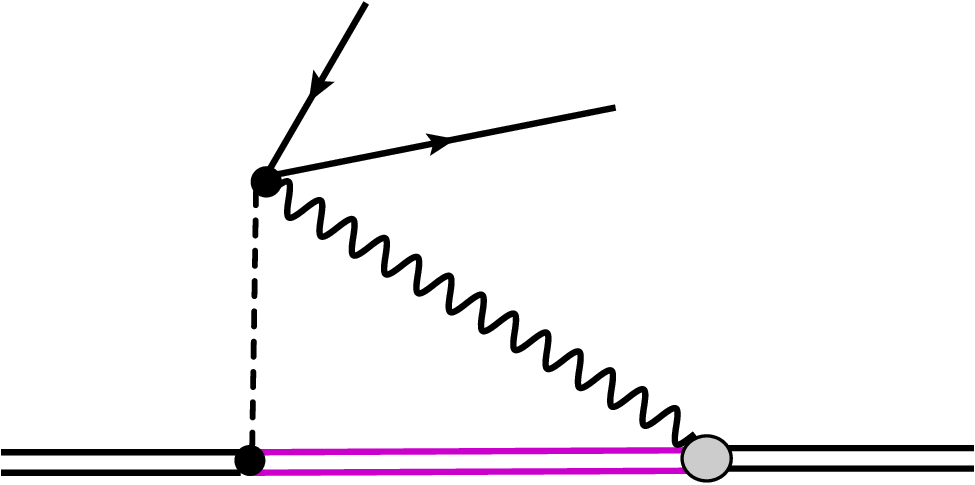}
         \caption{}
         \label{fig:trianglepion}
    \end{subfigure}
    \hspace{0.02\textwidth}
     \begin{subfigure}[b]{0.22\textwidth}
         \centering
         \includegraphics[width=\textwidth]{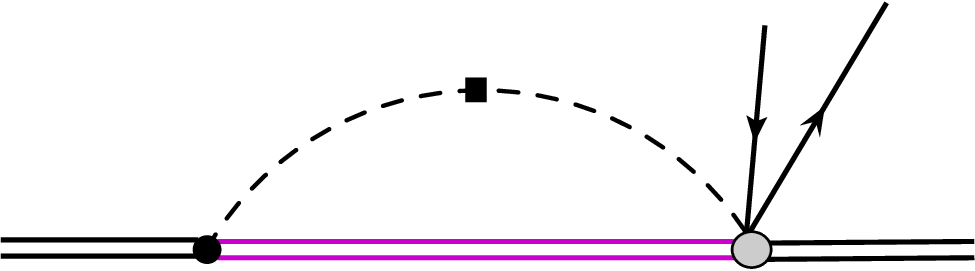}
         \caption{}
         \label{fig:NLOb4}
    \end{subfigure}
    \caption{LO and  NLO  diagrams contributing to the neutron decay, with a $\Delta$ intermediate state. We denote the $\Delta$ with the double pink lines. Single, double, dashed, and wavy lines denote leptons, nucleons, pions and photons, respectively. Dots refer to the interactions coming from the LO baryon Lagrangian $\mathcal{L}^{(1)}_{\pi N (\Delta)}$, squares denote the electromagnetic insertions $\mathcal{L}^{e^2 p^0}_{\pi}$, and grey circles denote interactions from the NLO Lagrangian $\mathcal{L}^{(2)}_{ N \Delta}$.}
    \label{fig:pion loops}
\end{figure*}

We compute the diagrams contributing to neutron $\beta$ decay, with intermediate $\Delta$ states, at order $\mathcal{O}(\alpha\epsilon_\chi^0)$ and $\mathcal{O}(\alpha\epsilon_\chi^1)$, which are shown in Fig. \ref{fig:pion loops}.
Using the power counting scheme presented in the previous section, we estimate the diagrams to be 
\begin{equation}\label{eq:PCestimates}
    \begin{split}
        \mathcal{A}_{a,b,c}&\sim \frac{\alpha}{4\pi}G_F h^2_A \sim \mathcal O(\epsilon_\chi^0)\times \alpha G_F\\\mathcal{A}_{d,e}&\sim \frac{\alpha}{4\pi}G_F h_A g_A \sim \mathcal O(\epsilon_\chi^0)\times \alpha G_F
        \\
        \mathcal{A}_{f,g,h}&\sim \frac{\alpha}{4\pi}G_F h_A b_1 \frac{Q}{m_N} \sim \mathcal O(\epsilon_\chi^1)\times \alpha G_F \\
        \mathcal{A}_{i}&\sim \frac{\alpha}{4\pi}G_F h_A b_{4,5} Q \sim \mathcal O(\epsilon_\chi^1)\times\alpha \;G_F
    \end{split}
\end{equation}
where $h_A,g_A,b_1$ are dimensionless order one coefficients and $b_{4,5}\sim\mathcal O(1/\Lambda_
\chi)$ are dimensionful coefficients.

Explicit calculation shows that the diagram ~\ref{fig:pion3} vanishes at LO. This is because of the tensor structure of the loop integrals: after integration, the only available Lorentz vector is the velocity $v_\mu$, and contracting it with the spin-$3/2$ projector gives zero.

Diagrams ~\ref{fig:pion1}, \ref{fig:pion2}, \ref{fig:pion4} and \ref{fig:pion5} give contributions both to the vector and axial operators. As expected from the conservation of the vector current, the vector contribution cancels in the $q^2\rightarrow 0$ limit after including wavefunction renormalization (WFR). Only an axial correction remains and the resulting leading-order amplitude can be written as
\begin{eqnarray}\label{eq:LOamplitude}
&&\mathcal{A}_{\text{LO}}=\frac{\alpha}{2\pi}Z_\pi\;\bar l_\rho S^\rho\, g_A h_A^2
\Bigg[\frac{16}{3}\left(\frac{m_\pi^2}{\Delta^2}-\frac{5}{18}\frac{g_1}{g_A}\right)f(\Delta,m_\pi)\;\;\nonumber\\
&&-\frac{8 \pi m_\pi}{3\Delta}+\frac{64}{9}-\frac{124}{81}\frac{g_1}{g_A}+\frac{20}{27}\frac{g_1}{g_A}\left(\frac{1}{\hat\epsilon}+\log\frac{\mu^2}{m_\pi^2}\right)\Bigg]\,.
\end{eqnarray}
where the loop function is defined as
\beq
f(\Delta,m_\pi)=\frac{
\Delta }{  (\Delta^2-m_\pi^2)^{1/2}}\log\left(\frac{\Delta + \sqrt{ \Delta^2-m_\pi^2 }}{m_\pi}\right)\,,
\eeq
and we defined
\begin{equation}
   \frac{1}{\hat\epsilon }=\frac{1}{\epsilon}-\gamma_E+\ln 4\pi\,.
\end{equation}
More details on the explicit calculation can be found in Appendix \ref{Appendix}. The result in Eq.~\eqref{eq:LOamplitude} contains three distinct contributions: (i) ultraviolet (UV) divergences, proportional to $1/\hat{\epsilon}$, (ii) decoupling-breaking terms (DBTs) that arise through the loop function and grow with the $\Delta$-nucleon mass splitting, and (iii) finite low-energy contributions. 

The UV divergences are canceled by local operators in the $\pi N$ effective Lagrangian  \cite{Gasser:2002am, Cirigliano:2022hob}. 
The appearance of terms proportional to positive powers of $\Delta$ does not indicate a breakdown of the power counting. Rather, this reflects that the loop integrals are sensitive to momentum regions where $\Delta$ effectively behaves as a large scale. At low energies, such contributions act as local terms and can be absorbed into the LECs of the effective theory.

As discussed above, we follow Ref.~\cite{Siemens:2017iuh} to identify the DBTs using the method of regions. In the limit $\Delta\gg m_\pi$, we expand the loop integrand in the hard region where the loop momentum scales as $k\sim \Delta$, while treating $\Delta$ as a heavy scale and the pion mass and external momenta as small scales. We only keep the leading term of this expansion which is independent of the light scales and has the same operator structure as the leading local counterterm. Higher-order terms in the hard expansion are proportional to additional powers of $m_\pi$ and correspond to higher-order operators in the EFT, so we do not subtract them. Denoting this leading hard contribution by $\mathcal A_{\rm hard}$, we define the renormalized amplitude as
\begin{equation}
\mathcal A_{\rm ren}= \mathcal A_{\rm loop}-\mathcal A_{\rm hard}+ \mathcal A_{\rm ct}\;,
\end{equation}
where $\mathcal A_{\rm ct}$ is the counterterm contribution. By construction, the subtraction removes the DBTs from the loop amplitude, while the counterterm absorbs the remaining ultraviolet divergence together with the leading local contribution from the hard region. This is analogous to extended-on-mass-shell (EOMS) subtractions, in which power-counting-breaking pieces are absorbed into LECs ~\cite{Gegelia:1999gf,Fuchs:2003qc}. A detailed example of this subtraction scheme is presented in Appendix~\ref{Appendix}.

Now we are ready to match the renormalized amplitude to the pionless theory, and find the shift to the axial coupling to be
\begin{eqnarray}\label{eq:ShiftLODelta}
     \Delta_{A, \rm em}^{\Delta{(0)}}&=&
 Z_\pi h_A^2 \Bigg[\frac{40}{27}\frac{g_1}{g_A}\left(f(\Delta,m_\pi)-\log\frac{2\Delta}{m_\pi}\right)\nonumber\\&& - \frac{16}{3}\frac{m_\pi^2}{\Delta^2}f(\Delta,m_\pi) + \frac{8\pi}{3}\frac{m_\pi}{\Delta} \Bigg]+\hat{C}^{(0)}_A.
\end{eqnarray}
The corresponding counterterm $\hat C^{(0)}_A$ has the same operator structure as the axial current in the pionless theory and can therefore be absorbed into the renormalization of the effective axial coupling. We notice that the last term of the LO correction is enhanced by a factor of $8\pi/3$. This can also be understood from the $\Delta$ resonance saturation. Matching the $\Delta$-full theory onto the $\Delta$-less theory at tree level, gives $c_3^{(\Delta)}=-2c_4^{(\Delta)}=-\frac{4h_A^2}{9\Delta}$  \cite{Epelbaum:2007sq}. Inserting these into the NLO result given in Eq.~\ref{c4c3} we obtain 
\begin{equation}
     Z_\pi  4 \pi m_\pi (c_4 - c_3) \rightarrow Z_\pi  \frac{8 \pi m_\pi}{3} \frac{h_A^2}{\Delta}.
\end{equation} which reproduces exactly the last term in Eq.~\eqref{eq:ShiftLODelta}.\footnote{We thank Martin Hoferichter for pointing this out.}

Following a similar analysis, we compute the NLO amplitude, which receives contributions from the diagrams shown in the lower part of Fig.~\ref{fig:pion loops}. 
Specifically, diagram ~\ref{fig:vertex1} gives a contribution to the axial operator and is proportional to $b_1$, as it involves the magnetic dipole transition $\
N\Delta\gamma$. At this order, the remaining corrections arise from the subleading interactions $ N\Delta\pi$ proportional to $b_4$ and $b_5$ through the different topologies of diagram ~\ref{fig:NLOb4}. These corrections turn out to be small, and we have neglected corrections from couplings fixed by reparametrization invariance. Putting everything together, we obtain
 \begin{align}\label{eq:shiftNLOdelta}
&\Delta_{A, \rm em}^{\Delta{(1)}}= \; Z_\pi \frac{h_A}{g_A}\Delta \Bigg[ \frac{  2 b_1 }{9\;  m_N Z_\pi}\left(\log{\frac{\mu^2}{4\Delta^2} }-\frac{4}{3}\right)+\hat C^{(1)}_A(\mu) \\
& -\frac{4}{9}\left( 18 b_4 + {17} b_5\right) \left( \frac{\Delta^2 - m_\pi^2}{\Delta^2} f(\Delta,m_\pi) - \log \frac{2\Delta}{m_\pi} \right)\Bigg]
.\nonumber
\end{align}
The scale dependence in Eq.~\eqref{eq:shiftNLOdelta} is absorbed by a local counterterm which has the same chiral structure as $\hat C_A$, for which we can use the expansion in Eq.~\eqref{eq:ctexp}.
Finally, we note that terms of the form $\Delta/m_N$ vanish and do not require additional subtraction, because $\Delta/m_N\rightarrow0$ in the decoupling limit by definition in the SSE \cite{Siemens:2020vop}.

Finally, we identify a new, finite contribution to weak magnetism through diagram \ref{fig:pion1}
\begin{equation}\label{eq: magnetic correction}
    \mu^\Delta_{\rm weak}=\frac{\alpha}{2\pi} \frac{
Z_\pi h_A^2 m_N }{  9(\Delta^2-m_\pi^2)^{1/2}}\log\left [\frac{\Delta + \sqrt{\Delta^2-m_\pi^2 }}{\Delta - \sqrt{ \Delta^2-m_\pi^2 }}\right ],
\end{equation}
which in the SSE appears at the same order as Eq.~\eqref{eq:WMPRL}. The presence of $m_N$ in the numerator arises from the definition of $\mu_{
\mathrm{weak}}$ in Eq.~\eqref{eq:Lpionless} and the result is safe in the decoupling limit.  There are also new contributions from diagrams \ref{fig:box} and \ref{fig:trianglepion} to the tensor coupling, but are suppressed by $\epsilon_\chi$ compared to Eq.~\eqref{eq:cTPRL}, which is already too small to be detected. We therefore do not give the $c_T^\Delta$ corrections here. 

\section{Numerical impact}\label{sec:Numerics}
We now compare our results to those of Ref.~\cite{Cirigliano:2022hob}, where the $\Delta$ resonance was not explicitly included, but its effect was incorporated through the low-energy coefficients $c_3,c_4$ that appear in the shift of the axial coupling at NLO. Thus, within the $\Delta$-full EFT, we combine the corrections computed in \cite{Cirigliano:2022hob} with the new corrections from the $\Delta$-full theory, presented in this work. The radiative correction to the order we are working is then computed as
\begin{equation}\label{eq: general RC}
\delta_\mathrm{RC} =
\frac{\alpha}{2 \pi} \left( \Delta^{(0)}_{A, \rm em}  + \Delta^{(1)}_{A, \rm em}+\Delta_{A, \rm em}^{\Delta{(0)}} + \Delta_{A,\rm em}^{\Delta{(1)}} - \Delta^{(0)}_{V,\rm em}
\right).
\end{equation}
The first two terms correspond to the pion-induced corrections of Ref. \cite{Cirigliano:2022hob}, while $\Delta^{\Delta{(0)}}_{A,\rm em}$ and $\Delta^{\Delta{(1)}}_{A,\rm em}$ represent the new contributions arising from explicit $\Delta$ degrees of freedom. 

For the numerical analysis, we use $Z_\pi = 0.81$, obtained from the physical pion mass splitting, and the average nucleon mass $m_N = 938.9$ MeV. In the loop amplitudes $g_A$ is well determined, with $g_A = 1.27$, while for $h_A,g_1$, we use the large $N_c$ relations~\cite{Siemens:2016jwj}, which give $h_A=1.40 \pm 0.05\;, \;\; g_1=2.32\pm0.26$. The NLO low-energy constants $c_3$ and $c_4$ are taken from the pion-nucleon scattering analysis of Ref.~\cite{Siemens:2016jwj}, where values are provided both in the $\Delta$-less and $\Delta$-full formulations of the EFT. The value of $b_1$ can be determined from the $\Delta\rightarrow N\gamma$ decay width, we adopt the commonly used value of $b_1\simeq3.8$ ~\cite{McGovern:2012ew,Pascalutsa:2005vq}. The individual values of subleading $\pi N\Delta$ LECs $b_4$ and $b_5$ are not well determined. We use $b_4=- b_5\simeq 1.3$ GeV$^{-1}$, values derived from the combinations $b_4\pm b_5$ in terms of the fitted $\pi N $ LECs, obtained from the large $N_c$ relations ~\cite{Siemens:2016jwj}.

It is useful to first look at the contribution computed in the theory without explicit $\Delta$. Using the fit values obtained in the $\Delta$-less theory ~\cite{Hoferichter:2015tha, Siemens:2016jwj},
\begin{eqnarray}
c_3 &=&-\{3.61(5),\,5.39(5),\,5.67(6)\}\, {\rm GeV}^{-1} \,,\nonumber \\
c_4 &=&\phantom{-}\{2.17(3),\,3.62(3),\,4.35(4)\}\, {\rm GeV}^{-1}\,,
\end{eqnarray}
denoting NLO, N$^2$LO, and N$^3$LO extractions, the corrections from the pion-induced loops are
\begin{equation}\label{pion_corrections}
\Delta^{(0)}_{A-V, \rm em} \in \{ 2.4, \, 5.7 \}\,, \
 \Delta^{(1)}_{A, \rm em} = \{ 10.0,    14.5,    15.9 \}.
\end{equation}
Here $\Delta^{(0)}_{A-V, \rm em}$ is obtained by setting $\hat C_{A}(\mu) - \hat{C}_V=0$ and varying $\mu$ between $0.5$ and $1$ GeV. The variation of the scale should be interpreted as an estimate of the size of the unknown short-distance contributions.

In the $\Delta$-full EFT, the same pion-loop corrections are present, but we use the modified values of the LECs \cite{Siemens:2016jwj,Siemens:2020vop} \footnote{We note that the presented values are given in \cite{Siemens:2020vop}, and slightly differ from \cite{Siemens:2016jwj}, however, this does not have a numerical impact here.}
\begin{eqnarray}
c_3^\Delta &=&-\{0.44(23),\,2.75(84),\,2.04(39)\}\, {\rm GeV}^{-1} \,,\nonumber \\
c_4^\Delta &=&\phantom{-}\{0.64(11),\,1.58(16),\,2.07(29) \}\, {\rm GeV}^{-1}\,,
\end{eqnarray}
resulting in notably smaller corrections
\begin{equation}\label{deltapion_corr}
 \Delta'^{(1)}_{A, \rm em} = \{ 3.4,   7.8 ,    7.5 \},
\end{equation}
where we used the prime to distinguish the use of the fit values obtained in the $\Delta$-full theory. While reduced, the NLO corrections are still larger than the LO diagrams and the convergence of the series, already noticed in Ref.~\cite{Cirigliano:2022hob}, remains worrisome. 

The fact that including the $\Delta$ leads to smaller values for pion loop diagrams is expected because in the $\Delta$-less theory the low-energy constants $c_3$ and $c_4$ receive sizeable contributions from integrating out the $\Delta$ resonance. Once the $\Delta$ is included explicitly, part of this contribution is removed from the LECs and appears instead through explicit $\Delta$-loop effects.
The natural question is then whether the contributions from the explicit $\Delta$ loops compensate for this reduction. For the numerical estimate of these loop corrections, we follow the same procedure by setting the counterterm contributions to zero and varying the renormalization scale $\mu$ between $0.5$ and $1$ GeV. This leads to  
\begin{equation}\label{delta_corrections}
\Delta^{\Delta(0)}_{A-V, \rm em} =3.8\,, \
 \Delta^{\Delta(1)}_{A, \rm em} \in - \{ 0.05,\   0.45 \},
\end{equation}
where the LO contribution is significantly larger than the NLO.  The LO contribution in Eq. \eqref{eq:ShiftLODelta} contains a term proportional to $m_\pi/\Delta\sim\mathcal O(1)$ which is enhanced by a factor of $\pi$, resulting in its relatively large size.  

We now make a comparison between the $\Delta$-less and the $\Delta$-full results. As discussed above, including the $\Delta$ leads to smaller values for $c_3,\;c_4$, and consequently reduces the NLO pion contribution from Eq.~\eqref{pion_corrections} to ~\eqref{deltapion_corr}. The reduction of the NLO contributions is partially compensated in the $\Delta$ full theory by LO $\Delta$ loops that are dominated by the resonance saturation pieces. This reshuffling significantly improves the convergence of the chiral expansion of the $\Delta$-full theory compared to the $\Delta$-less theory. In particular, using the NLO extractions for $c^\Delta_{3,4}$, we obtain 
\begin{equation}
\delta_\mathrm{RC}(\Delta\mathrm{-full}) = \frac{\alpha}{2\pi}\left[7.8(1.7) + 3.15(0.20)\right]\,,
\end{equation}
where the terms in square brackets are respectively the LO and NLO results in the $\Delta$-full theory. Using N${}^2$LO or N${}^3$LO values for the LECs worsens the convergence, but at this level other contributions for $\delta_{\mathrm{RC}}$ should be considered.  

We now compute the radiative correction to the electromagnetic shift $\lambda=g_A/g_V$, given in Eq.~\eqref{eq: general RC} also including the range from using higher-order $c^\Delta_{3,4}$ values and obtain
\begin{eqnarray}
    \delta^{\Delta(\lambda)}_\mathrm{RC}
\in   \{ 1.1, \,2.0 \} \cdot 10^{-2}\,.
\end{eqnarray}
The radiative correction in an EFT with the explicit inclusion of the $\Delta$ resonance is thus found to be moderately smaller than that presented in \cite{Cirigliano:2022hob}, $\delta^{(\lambda)}_\mathrm{RC}
\in   \{ 1.4, \,2.6 \} \cdot 10^{-2}\,$,
{but} it remains at the percent level. The main conclusion of Ref.~\cite{Cirigliano:2022hob} that the electromagnetic radiative corrections are relevant for the comparison between the neutron decay experiment and lattice QCD still holds.

The correction has no impact on the current first-row CKM unitarity tension, since the most precise determination of $\lambda=g_A/g_V$ comes from neutron decay experiments, where these radiative effects are already included. That being said, the experimental value of  $\lambda$ extracted from neutron decay is related to $g_A$ in pure QCD, through the equation ~\cite{Bhattacharya:2011qm}
\begin{equation} \label{eq:exp_to_lattice}
    \lambda = g_A^\mathrm{QCD} \Big(1 +\delta^{(\lambda)}_\mathrm{RC}  - \epsilon_R \Big) ,
\end{equation}
where $\delta^{(\lambda)}_\mathrm{RC}$ is the radiative correction in Eq.~\eqref{eq: general RC} and $\epsilon_R$ parametrizes a possible BSM right-handed (RH) charged-current interaction.

The latest lattice calculation of $g_A$ in the isospin limit, was performed in \cite{Hall:2025ytt} with sub-percent precision,
\begin{equation}
    g_A^{\rm QCD}= 1.2674(96).
\end{equation}
Since present lattice calculations are done in pure QCD and neglect electromagnetic effects, Eq.~\eqref{eq:exp_to_lattice}  should be used for the comparison to the experimental measurement.
\begin{figure} [t]
    \centering
    \includegraphics[width=1.0\linewidth]{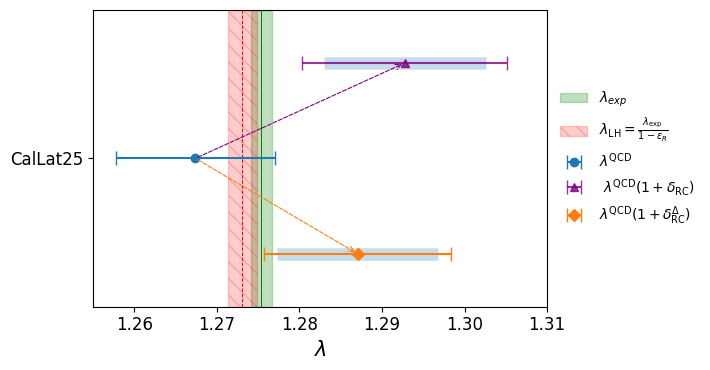}
    \caption{Impact of the required electromagnetic shift of the lattice QCD determination of $g_A$ from CalLat25 \cite{Hall:2025ytt} in comparison with the experimentally extracted value of $\lambda$. Blue circles denote pure-QCD lattice results, orange diamonds show the values obtained after including the radiative correction calculated in this work, purple triangles indicate the shift corresponding to the previous estimate from \cite{Cirigliano:2022hob}. The green band indicates the experimental determination $\lambda=1.2754(13)$. The red hatched band shows the value preferred by a right-handed-current solution of the Cabibbo-angle anomaly, corresponding to $\epsilon_R=v^2[C_{Hud}]_{11}/V_{ud}$ \cite{Cirigliano:2023nol}.} 
    \label{fig:RCshift_CalLat25}
\end{figure}
In Fig. \ref{fig:RCshift_CalLat25} we show the role of electromagnetic corrections in this comparison. The blue point corresponds to the pure-QCD lattice result, while the green vertical band indicates the experimentally extracted value of $\lambda$. Applying the radiative correction calculated in this work shifts the lattice result to larger values. The radiative correction of Ref.~\cite{Cirigliano:2022hob} somewhat overshoots the experimental value, but this is not true with the reduced correction found in this work. While reduced, the shift is still sizeable compared to the experimental uncertainty.

 Assuming the best model average value for the RH coupling, to explain the CAA \cite{Cirigliano:2022yyo,Cirigliano:2023nol}, we find that $g_A$ would shift by around $0.2\%$. We visualize this shifted value with the red hatched band shown in Fig. \ref{fig:RCshift_CalLat25}. We observe that the shift induced by such a RH-current contribution is smaller than both the present lattice uncertainty and the electromagnetic correction. This indicates how important it is to improve theoretical calculations and reduce the errors of $g_A^{\rm QCD}$ and $\delta_{\rm RC}$, in order to be able to make precision comparisons between experiment, lattice-QCD, and BSM interpretations.

 Finally, we compute the impact of $\mu_{\mathrm{weak}}^\Delta$ and find the correction to be an order of magnitude smaller than Eq.~\eqref{eq:WMPRL}, rendering the $\Delta$ contributions negligible in light of expected experimental sensitivities.

\section{Discussion}\label{sec:discussion}
In this work, we studied electromagnetic radiative corrections to neutron $\beta$ decay within $\chi$PT with explicit $\Delta$ degrees of freedom. Following the analysis of Ref.~\cite{Cirigliano:2022hob}, we identified and computed additional contributions arising from diagrams with intermediate $\Delta$ states up to NLO. After performing the matching to the pionless EFT, we found the corresponding electromagnetic shifts to the axial coupling, as well as additional isospin-breaking corrections to the weak magnetism term.

Including the $\Delta$ resonance affects the numerical estimate of the radiative correction mostly in two ways. Firstly, it generates additional loop contributions to the neutron decay. Secondly, it changes the values of the LECs $c_3$ and $c_4$ that enter the dominant correction from the NLO pion-loops. Using the $\Delta$-full values of these couplings, and combining them with the explicit $\Delta$ contributions calculated in this work, we obtain a correction that is about  $25\%$ smaller than in the $\Delta$-less theory. We find that the convergence of the chiral expansion is much improved after including the $\Delta$ as an explicit degree of freedom. 

Even though the size of the correction is reduced with the inclusion of the $\Delta$ resonance, the effect remains at the percent level, and is therefore very relevant in the comparison between lattice QCD calculations of $g_A/g_V$ and experimental extractions. In particular, the uncertainty on the radiative corrections is at the $0.9\%$ level and about four times larger than possible BSM effects that can explain the Cabibbo-angle anomaly. 

As in the case of the $\Delta$-less calculation, a determination of the LEC $\hat{C}_A$ is crucial to fully assess the size of radiative corrections to $g_A$.
The non-perturbative correlation functions to be calculated in lattice QCD and in lattice QCD+QED 
, and strategies to match them to $\chi$PT
have been identified in Refs. \cite{Seng:2024ker,Cirigliano:2024nfi}, where the matching relations were derived in a $\Delta$-less theory. It will be important to extend these calculations to include explicit $\Delta$ degrees of freedom, which will likely lead to a better converged chiral expansion.

\textbf{Note added:} Right after submitting the first version of this work, the authors of Ref. \cite{Tomalak:2026ido} informed us of
their own independent calculation of $\Delta$ induced radiative corrections. After careful comparison and discussion, we found mistakes in both our original calculations. The updated result in this v2 now agrees.

\begin{acknowledgments}
We thank Martin Hoferichter and André Walker-Loud for useful discussions. We also thank Oleksandr Tomalak for the detailed comparison of results.
LG is supported by the Dutch Research Council (NWO) in form of an M1 grant. JdV is funded by the Dutch Research Council (NWO) in the form of a VIDI grant and by the European Union (ERC, CRUNS, 101230525). Views and opinions expressed are, however,  those of the author(s) only and do not necessarily reflect those of the European Union or the European Research Council. Neither the European Union nor the granting authority can be held responsible for them.
E.M. is supported by the U.S. Department of Energy Office and by the Laboratory Directed Research and Development (LDRD) program of Los Alamos National Laboratory under project numbers 20250164ER and 20260246ER. Los Alamos National Laboratory is operated by Triad National Security, LLC, for the National Nuclear Security Administration of the U.S. Department of Energy (Contract No. 89233218CNA000001)

\end{acknowledgments}

\appendix
\section{Example of the decoupling scheme} \label{Appendix}
In this appendix we illustrate the decoupling scheme we used throughout the work for loop diagrams with a $\Delta(1232)$ intermediate state. As an example loop, we use diagram (b) of Fig.~\ref{fig:pion loops} and show the details of both the full calculation and the subtraction of the decoupling breaking terms that appear. We note at this point that this individual diagram is not a physical quantity, but this example goes through all the steps needed to implement the subtraction scheme.  We explain afterwards how the same procedure is applied to the rest of the loops that enter the neutron decay amplitude.

\subsection{Notation and master integrals}

Throughout this appendix, we work in dimensional regularization with
$d=4-2\epsilon$ and define
\begin{equation}
    \int_k \equiv \mu^{2\epsilon} \int\frac{d^d k}{(2\pi)^d}.
\end{equation}
For convenience, we introduce the leptonic charged current
\begin{equation}
    \bar l_\mu = -2\sqrt{2}\,G_F V_{ud}\, \bar u(p_e)\gamma_\mu P_L v(p_\nu).
\end{equation}
The loop integrals encountered in the pion-loop diagrams can be written in the general form
\begin{equation}
    I_{n,r}[N(k)] =\int_k\frac{N(k)}{(k^2-m_\pi^2+i0)^n (v\cdot k-\Delta+i0)^r},
\end{equation}
where $\Delta=m_\Delta-m_N$ is the nucleon-$\Delta$ mass splitting, and $N(k)$ denotes the dependence of the numerator on the loop momentum $k$. Since the heavy-baryon velocity $v^\mu$ is the only external four-vector appearing in the denominators, tensor integrals can be decomposed according to
\begin{equation} \label{decomposition}
\begin{split}
I_{n,r}[k_{\mu}k_\nu]\rightarrow &\;\;A_{n,r}\; g_{\mu\nu}  +B_{n,r}\; v_\mu
v_\nu\\
     I_{n,r}[k_{\mu}k_\nu k_\rho]\rightarrow &\;\;C_{n,r}\; v_\mu
v_\nu v_\rho \\&+D_{n,r}\left(g_{\mu\nu}v_\rho+g_{\mu\rho}v_\nu+g_{\rho\nu}v_\mu \right)
\end{split}
\end{equation}
where after contracting these expressions with $g_{\mu\nu}$ and $v_\mu$, we can find
\begin{equation}
\begin{split}
     A_{n,r}&=\frac{1}{d-1} I_{n,r}\left[k^2 - (v\cdot k)^2\right]\\D_{n,r}&=\frac{1}{d-1} I_{n,r}\left[k^2 v\cdot k- (v\cdot k)^3\right]
    \end{split}
\end{equation}
We can then reduce the powers of the loop momentum in the numerator using the usual identities repeatedly, until all tensor integrals are reduced to scalar loop functions. When $r=0$ we get standard tadpole-like integrals like
\beq 
I_{2,0}[1]=\int_{k} \frac{1}{\left(k^2-m_\pi^2\right)^2} =\frac{i}{16 \pi^2}\left(\frac{1}{\hat\epsilon}+\log{\frac{\mu^2}{m_\pi^2}}\right)\;,
\eeq
where we introduced the short notation
\begin{equation}
   \frac{1}{\hat\epsilon }\equiv\frac{1}{\epsilon}-\gamma_E+\log 4\pi 
\end{equation}
The remaining scalar integrals can be generated from the two-point function
\begin{equation}
    I_{1,1}=\int_k \frac{1 }{\left(k^2-m_\pi^2\right)} \frac{1}{(v\cdot k-\Delta+i0)}
\end{equation}
Using the heavy-particle two-point function of
Ref.~\cite{Zupan:2002je}, one obtains, for $m_\pi<\Delta$,
\begin{equation}
    I_{1,1} =-\frac{2i\Delta}{(4\pi)^2} \left[   \frac{1}{\hat\epsilon} +\log{\frac{\mu^2}{m_\pi^2}} +2 -2F\left(\frac{m_\pi}{\Delta}\right) \right],
\end{equation}
with
\begin{equation}
    F\left(\frac{m_\pi}{\Delta}\right)= \frac{\sqrt{\Delta^2-m_\pi^2}}{\Delta}\log\left( \frac{\Delta+\sqrt{\Delta^2-m_\pi^2}}{m_\pi}\right).
\end{equation}
The scalar integrals  with $n,r\geq1$ are generated from the master integral $I_{1,1}$ by differentiating with respect to $m_\pi^2$ and $\Delta$.

\subsection{Example loop : Diagram (b)}
We show here how the decoupling scheme is applied in diagram \ref{fig:pion2}. Since we are interested in the corrections to the vector and axial charges, we evaluate the amplitude assuming that the external momentum transfer goes to zero. After summing the two allowed charge topologies, the amplitude is
\begin{eqnarray}
\label{eq:Ab}
    &\mathcal{A}_{\text{b}}=i \frac{h_A^2\delta m_\pi^2}{F^2} \bar l_\rho\left[(2-d)v^\rho  +\frac{2}{3}\left[(d-3)+\frac{(d-3)^2}{(d-1)^2}\right]g_1S^\rho\right]\nonumber\\& \times\frac{1}{d-1}\Big[I_{1,2}-I_{2,0} -2\Delta I_{2,1}+  \left(m^2_\pi-\Delta^2\right)I_ {2,2} \Big]   
\end{eqnarray}
where we used the tensor decomposition introduced above, together with $P^{\mu\nu}_{3/2}v_\nu=0$. Evaluating the scalar integrals gives
\begin{eqnarray}\label{eq:diagbfull}
&\mathcal{A}_{\text{b}}^{\text{loop}}=\frac{\alpha}{\pi}Z_\pi h_A^2\,\bar l_\rho \Bigg [v^\rho \;\left(1-\frac{1}{\hat\epsilon} -\log{\frac{\mu^2}{m_\pi^2}} +2 f(\Delta,m_\pi) \right)\nonumber\\&-\frac{10}{27}g_1S^\rho\left(\frac{31}{15}-\frac{1}{\hat\epsilon}-\log\frac{\mu^2}{m_\pi^2}+2f(\Delta,m_\pi)\right)\Bigg] 
\end{eqnarray}
where the loop function is defined as
\beq f(\Delta,m_\pi)=\frac{\Delta }{  \sqrt{\Delta^2-m_\pi^2}}\log\left(\frac{\Delta + \sqrt{ \Delta^2-m_\pi^2 }}{m_\pi}\right)\,.
\eeq
Eq.~\eqref{eq:Ab} is the full one-loop contribution, besides the ultraviolet divergence, it contains the loop function $f(\Delta,m_\pi)$, which depends on both the pion mass and the $\Delta$-nucleon mass splitting. In the large $\Delta$ limit, the loop function behaves for $\Delta\gg m_\pi$ as
\begin{equation}
    f(\Delta,m_\pi)=\log\frac{2\Delta}{m_\pi}+ \frac{m_\pi^2}{2\Delta^2}\left(\log\frac{2\Delta}{m_\pi}-\frac{1}{2}\right)+\dots
    \label{eq:f-largeDelta}
\end{equation}
where the first term is logarithmic in $\Delta$ and blows up in the decoupling limit $\Delta\rightarrow \infty$. Eq.~\eqref{eq:diagbfull} therefore contains terms that do not vanish in the limit $\Delta\to\infty$, and must be absorbed into local operators. These are precisely the decoupling breaking terms (DBTs) discussed in Sec.~\ref{sec:deltaloops}.

To identify these terms, we can seperate the integral using the method of regions. In the large $\Delta$ limit, the loop receives contributions from two momentum regions:
\begin{itemize}
    \item soft momentum region, $k\sim m_\pi$,
    \item hard momentum region, $k\sim \Delta$.
\end{itemize}
Hence, following the procedure introduced in Sec.~\ref{sec:deltaloops}, we define the hard loop by expanding the integrand in powers of the small parameters before integrating, treating the loop momentum and $\Delta$ as large scales. In the regime $m_\pi \ll k$, we expand
\begin{equation}
\frac{1}{k^2 - m_\pi^2} = \frac{1}{k^2} + \mathcal{O}\left(\frac{m_\pi^2}{k^2}\right) \, ,
\end{equation}
so that at leading order the integrals reduce to
\begin{equation}
I_{n,r}^{\text{hard}} \sim \int_k \frac{1}{(k^2)^n (v\cdot k - \Delta)^r} \, .
\end{equation}
Note that $I_{n,0}^{\text{hard}} $ vanishes in dimensional regularization, since these integrals are scaleless. Therefore, only integrals with $r \geq 1$ survive. Useful standard integrals here are 
\begin{equation}
\int d^dk \left(\frac{-1}{k^2}\right)^\alpha \left(\frac{\omega}{\omega+v\cdot k}\right)^p = i \pi^{d/2}(-2\omega)^{d-2\alpha} I(\alpha,p)\,,
\end{equation}
where 
\begin{equation}
I(\alpha,p) = \frac{\Gamma(2\alpha+p - d)\Gamma(d/2-\alpha)}{\Gamma(\alpha)\Gamma(p)}\,.
\end{equation}
Keeping only the leading term of the hard expansion, which is independent of the scale $m_\pi$, we get
\begin{eqnarray}\label{eq:exploop_b}
\mathcal{A}^{\text{hard}}_{\text{b}}&&=\frac{\alpha}{\pi}Z_\pi h_A^2\,\bar l_\rho \Bigg [v^\rho \;\left(1-\frac{1}{\hat\epsilon}  +2\log\frac{2\Delta}{\mu}  
\right )\nonumber\\&&-\frac{10}{27}g_1S^\rho\left(\frac{31}{15}-\frac{1}{\hat\epsilon}+2\log\frac{2\Delta}{\mu}\right)\Bigg] 
\end{eqnarray}
Here we observe, as expected, the same terms that break the decoupling that arise from expanding the loop function in Eq.~\eqref{eq:diagbfull}, as well as the UV divergences associated with the expanded integrals.  Higher-order terms in the hard expansion are proportional to powers of $m_\pi^2/\Delta^2$ and correspond to higher-order operators in the EFT. Since such operators are beyond the order we consider here, they are not included in the subtraction scheme.
We now define the renormalized amplitude as
\begin{equation}
\mathcal A^{\rm ren}= \mathcal A^{\rm loop}-\mathcal A^{\rm hard}+ \mathcal A^{\rm ct}\;,
\end{equation}
where $\mathcal A^{\rm ct}$ is the counterterm contribution. By construction, the subtraction removes the DBTs from the full loop amplitude, while the counterterm absorbs the UV divergences of both the full loop and the expanded contribution and contains finite pieces needed to absorb the DBTs. The corresponding counterterm has the same structure as the axial current in the pionless theory and can be absorbed into the renormalization of the effective axial coupling. We can express the bare coupling at leading order in the SSE expansion as
\begin{eqnarray}\label{eq:CArenorm}
    \hat C_{A,\mathrm{bare}}^{(0)}=\hat C_A^{(0)}(\mu)+\delta\hat C_A^{(0)}+\delta\hat C_A^{(\Delta,0)},
\end{eqnarray}
where $\hat C_A^{(0)}(\mu)$ is the renormalized coefficient, $\delta\hat C_A^{(0)}$ removes the UV divergence from the pion loops, and $\delta\hat C_A^{(\Delta,0)}$ absorbs the additional UV divergences together with the finite DBTs arising from the subtraction scheme.

Combining the full loop expression with the hard loop and the counterterm contribution, the renormalized amplitude of the topology \ref{fig:pion2} is 
\begin{eqnarray}\label{eq:renormloop_b}
\mathcal{A}^{\text{ren}}_{\text{b}}&=&4\left(\frac{ \alpha}{2\pi } Z_\pi\right)  h_A^2\;\;\bar l_\rho \left(v^\rho  -\frac{10}{27}g_1S^\rho \right)\\&\times&\left(f(\Delta,m_\pi)-\log\frac{2\Delta}{m_\pi}\right) +\hat C_A^{(0)},\nonumber
\end{eqnarray}
which is finite and behaves well in the decoupling limit. We stress that this example is shown here only to illustrate the subtraction scheme and the renormalization procedure for a single loop diagram. The same scheme is then applied to each loop diagram that contributes at a given order in the SSE expansion, and the physical amplitudes presented in Sec.~\ref{sec:deltaloops} at LO and NLO are obtained by summing all renormalized diagrams and counterterm contributions. This procedure guarantees that the final results are finite and obey the decoupling theorem.

The counterterms appearing in the LO and NLO amplitudes can be viewed as terms of the SSE expansion of the same local coefficient
\begin{equation}
 \hat{C}_A = \hat C_A^{(0)} + \frac{\Delta}{\Lambda_\chi} \hat C_A^{(1)} + \ldots, 
\end{equation}
and at each order the corresponding coefficient $\hat C_A^{(n)}$ is renormalized according to Eq.~\eqref{eq:CArenorm}.

\bibliography{references}

\end{document}